\documentclass[pss]{wiley2sp} % provides pss two-column style
\usepackage{amsmath}
\begin{document}

% Title of the article
\title{Quantum phase transition in a dimerized chain with hexamer distortion}

% Abbreviated title for the page headers
%\titlerunning{Short title }

% Authors
\author{%
  Mahboobeh Shahri Naseri\textsuperscript{\textsf{\bfseries 1,3}},
  George I. Japaridze\textsuperscript{\textsf{\bfseries 2}},
  Saeed Mahdavifar\textsuperscript{\Ast,\textsf{\bfseries 1}},
  and Saber Farjami Shayesteh\textsuperscript{\textsf{\bfseries 1}}}

% Abbreviated list of authors for the page headers
%\authorrunning{First author et al.}

%E-mail-address of corresponding author
\mail{e-mail
  \textsf{mahdavifar@guilan.ac.ir}}

% author's affiliations/addresses
\institute{%
\textsuperscript{1}\,Department of Physics,
University of Guilan, 41335-1914, Rasht, Iran\\
  \textsuperscript{2}\,College of Engineering, Ilia State University, Cholokashvili Ave.
3-5, 0162 Tbilisi, Georgia\\
  \textsuperscript{3}\,Department of Physics, Payame Noor
University,19395-3697, Tehran, Iran}
%  \textsuperscript{3}\,Third address}

%\received{XXXX, revised XXXX, accepted XXXX} % do not change, will be filled in by the publisher
%\published{XXXX} % do not change, will be filled in by the publisher

% Please select about four verbal keywords for your manuscript.

\abstract{ \abstcol{ We consider the dimerized spin-$\frac{1}{2}$
Heisenberg chain with spin hexameric distortion of the exchange
pattern and study the zero-temperature phase diagram in the
parameter space $(J_{1}, J_{2}, J_{3})$ by continuum-limit
bosonization approach and the exact diagonalization method. The
phase diagram is rich and has two gaped dimer phases.}{ We obtain
an estimate of the critical line separating the different gapped
dimer phases by the bosonization approach. The existence of the
transition line and the difference between dimer phases is checked
numerically. The behavior of the energy gap and the dimer order
parameter supports the exact location of the gapless line. }}

 \maketitle

\section{Introduction}

The critical gapless phases emerging on the border lines separating  two gapped phases in low-dimensional spin systems has been the subject of studies for decades. The spin
$S=1/2$ Heisenberg chain with dimerization and frustration, is a
primer and well studied example of a such system which shows a
gapless phase in the ground state phase diagram
\cite{Chitra_Sen_95,BKJ_98,Jiang_01,Uhrig_04}. After the seminal
paper by Martin-Delgado et.al
\cite{Sierra_96}, great attention has been focused on
the studies of the same phenomena
in the other wide class of low-dimensional magnets
such as the spin ladders.
\cite{Sierra_98,Kotov_et_al_99,Cabra_Grynberg_99,WN_00,Okomoto_03,Sierra_07,Sierra_08,Chitov_08,Chitov_11}
Recently, the gapless phases on the border of different massive
phases has been discussed in two-dimensional spin systems
\cite{Kotov09,Sandvik_10,M_Vojta_11}.

 In this paper, we address this problem within the slightly different framework, namely we
study the stability of the
explicitly dimerized (gapped)
Heisenberg chain towards the perturbation caused by the different
commensurate modulation of the exchange pattern which higher
period itself causes a gapped state however breaks the favored by
the explicit dimer order.
%%%%%%%%%%%%%%%%%%%%%%%%%%%%%%%%%%%%%%%%%%%%%%
%%%%%%%%%%%%%%%%%%%%%%%%%%%%%%%%%%%%%%%%%%%%%%%%%%%%%%%%%%%%%%%
%%%%%%%%%%%%%%%%%%%%%%%%%%%%%%%%%%%%%%%%%%%%%%%%%%%%%%%%%%%%%%%%%%

%\section{The Model}

The Hamiltonian of the model that
under consideration  is defined as
%***********************************************************
\begin{eqnarray}
{\hat H}&=&{\hat H}_{0}+{\hat
 H}_{\lambda}\, ,
\label{Hamiltonian}
\end{eqnarray}
%***********************************************************
where
%***********************************************************
\begin{eqnarray}
{\hat H}_{0}=J\sum_{n=1}^{N}\left(1-(-1)^{n}\delta
\right)\textbf{S}_{n}\cdot\textbf{S}_{n+1},
 \end{eqnarray}
%***********************************************************
is the Hamiltonian of the dimerized chain with strong odd links
($\delta>0$) and
%***********************************************************
\begin{eqnarray}
{\hat H}_{\lambda}&=&\lambda J
\sum_{n=1}^{N}\,g(n)\textbf{S}_{n}\cdot\textbf{S}_{n+1} \, ,
\label{Hamiltonian-Hexamer-Chain}
\end{eqnarray}
%***********************************************************
where
\begin{eqnarray}
g(n) =1+\cos\pi n\
+2\left[\cos(\frac{\pi}{3}n)+\cos(\frac{2\pi}{3}n)\right] \, ,&
 \label{Jn}
\end{eqnarray}
%%%%%%%%%%%%%%%%%%%%%%%%%%%%%
%%%%%%%%%%%%%%%%%************FIG 1********************%%%%%%%%%%%%%%%%%%%%%%%%%%%%%%%%%%%%%%%%
\begin{figure}
\includegraphics[scale=1.0]{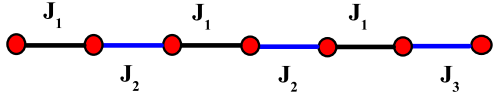}%,}%width=3.65in}}
 \caption{(Color online.) Schematic representation of
 spin chains with hexameric modulation
 of spin exchange which is considered in the paper.}
 \label{Schematic}
\end{figure}
%%%%%%%%%%%%%%%%%%%%%%%%%%%%%%%%%%%%%%%%%%%%%%%%%%%%%%%%%%
is the perturbation Hamiltonian of the dimerized chain with
commensurate and modulation of the exchange on each 6-th link
which increase (decrease) at $\lambda>0$ ($\lambda<0$). Here
$\textbf{S}_{n}$ is the spin $S=\frac{1}{2}$ operator at site $n$
and the chain consists of $N=6N_{0}$ sites. It is straightforward
to obtain the Hamiltonian Eq.
(\ref{Hamiltonian}), describes a
Heisenberg chain with the following hexamer modulation of the spin
exchange (see Fig.~\ref{Schematic}).
%%%%%%%%%%%%%%%%%%%%%%%%%%%%%%%%%%%%%%%%%%%%%%%%%%%%%%%%%%%%%%%%%%%%%%%%%%
%%%%%%%%%%%%%%
%***********************************************************
\begin{eqnarray}
J(1)&=&J(3)=J(5)=J(1+\delta)=J_{1}\nonumber\\
J(2)&=&J(4)=J(1-\delta)=J_{2},  \\
J(6)&=&J(1-\delta+6\lambda)=J_{3}.\nonumber
\end{eqnarray}
%***********************************************************

In our recent paper \cite{Paper_1}, we have studied the magnetic
phase diagram of the model in the limit of strong exchange on odd
links. It has been shown,  that the
presence of additional modulation
leads to the dynamical generation of two new energy scales in the
system and to the appearance of two additional quantum phase
transitions in the ground state of the system. These new gaps
appear at finite magnetization and show themselves in the presence
of two new magnetization plateaus at magnetization equal to
$\frac{1}{3}$ and $\frac{2}{3}$ of the saturation value.

In this paper, we study the effects caused by the additional
distortion of the exchange pattern with period six on the ground
state properties of the dimerized chain at zero magnetization. We
show that in the case of frustrating character of hexamer
distortion, with increasing $\lambda$ a quantum phase transition
takes place in the ground state of the system. The transition
corresponds to the reestablishment of the gapless critical phase
on the border line, which separates two dimerized phases with
shifted in respect to each other on one lattice unit dimer order.

The paper is organized as follows: In the next section, the
bosonization method is applied and the transition line is
determined analytically. In the section 3, the results of a
numerical experiment are presented. Finally, we
will discuss and summarize our
results in section $4$.

\section{Bosonization}

In this Section, we use the continuum-limit bosonization treatment
to study the ground state phase diagram of the model
(\ref{Hamiltonian}). To obtain the continuum version of the
Hamiltonian, we use the standard bosonization expression of the
spin operators \cite{Giamarchi_book_04}
%%%%%%%%%%%%%%%%%%%%%%%%%%%%%%%%%%%%%%%%%%%%%%%%%%%%%%%%%%%%%%%%
\begin{eqnarray}
 S_{n}^{z}  &=& \frac{1}{\sqrt{2\pi}}
\partial_x
\phi(x)+\frac{(-1)^{n}}{\pi a_{0}} \sin(\sqrt{2\pi}\phi)\, ,
\label{bosforTau_z}
\end{eqnarray}
%%%%%%%%%%%%%%%%%%%%%%%%%%%%%%%%%%%%%%%%%%%%%%%%%%%%%%%%%%%%%%%%%%
\begin{eqnarray}
S_{n}^{x} &=&\cos(\sqrt{2\pi}\theta)\left[1+ \frac{ (-1)^{n}}{\pi
a_{0}} \sin(\sqrt{2\pi}\phi)\right]\,
\textcolor[rgb]{0.98,0.00,0.00}{,}
\end{eqnarray}
%%%%%%%%%%%%%%%%%%%%%%%%%%%%%%%%%%%%%%%%%%%
\begin{eqnarray}
S_{n}^{y}&=&\sin(\sqrt{2\pi}\theta)\left[1-\frac{(-1)^{n}}{\pi
a_{0}}\sin(\sqrt{2\pi}\phi)\right]\, . \label{bosforTau_y}
\end{eqnarray}
%%%%%%%%%%%%%%%%%%%%%%%%%%%%%%%%%%%%%%%%%%%%%%%%%%%
%%%%%%%%%%%%%%%%%%%%%%%%%%%%%%%%%%%%%%%%%%%%%%%%%%%%%%%%%%%%%%%%%%%%%%%%%%%%%%%%%%%%%%%%%%%%%%%%%%%%%%%%%%
%%%%%%%%%%%%%%%%%%%%%%%%%%%%%%%%%%%%%%%%%%%%%%%%%%%%%%%%%%%%%%%%%%%%%%%%%%%%%%%%%%%%%%%%%%%%%%%%%%%%%%%%%%
%%%%%%%%%%%%%%%%%%%%%%%%%%%%%%%%%%%%%%%%%%%%%%%%%%%%%%%%%%%%%%%%%%%%%%%%%%%%%%%%%%%%%%%%%%%
%
Here $\phi(x)$ and $\theta(x)$ are dual bosonic fields,
$\partial_t \phi = v_{s}
\partial_x \theta $, and satisfy the following commutation
relations
%%%%%%%%%%%%%%%%%%%%%%%%%%%%%%%%%%%%%%%%%%%%%%%%%%%%%%%%%%%%%%%%
\begin{eqnarray}
\label{regcom}
&& [\phi(x),\theta(y)]  = i\Theta (y-x)\,,  \nonumber\\
&& [\phi(x),\theta(x)]  =i/2,
\end{eqnarray}
%%%%%%%%%%%%%%%%%%%%%%%%%%%%%%%%%%%%%%%%%%%%%%%%%%%%%%%%%%%%%%%%
and $a_{0}$ is the constant of the order of lattice unit. Using
(\ref{bosforTau_z})-(\ref{bosforTau_y}), one gets the following
bosonized expression for the alternating part of the "dimerization
operator"
%%%%%%%%%%%%%%%%%%%%%%%%%%%%%%%%%%%%%%%%%%%%%%%%%
\begin{eqnarray}
&& \textbf{S}_{n}\cdot\textbf{S}_{n+1} \sim
\frac{(-1)^{n}}{\pi^{2} a_{0}}\cos(\sqrt{2\pi}\phi(x)),
\label{bosforSnSn}
\end{eqnarray}
%%%%%%%%%%%%%%%%%%%%%%%%%%%%%%%%%%%%%%%%%%%%%%%%%%
and finally (see for details Ref. \cite{Paper_1}) the following
continuum-limit bosonized version of the Hamiltonian
(\ref{Hamiltonian})
%\begin{widetext}
%%%%%%%%%%%%%%%%%%%%%%%%%%%%%%%%%%%%%%%%%%%%%%%%%%%
\begin{eqnarray}
&&  H_{Bos}=\int dx \Big\{ \frac{u}{2}[(\partial_{x}\phi)^{2} +
(\partial_x\theta)^{2} ]\nonumber\\
&& \quad  - \, \frac{\Delta_{0}}{2\pi a_{0}} \cos\left(\sqrt{2\pi
}\phi\right)\Big\}, \label{Bosonized_Hamiltonian}
\end{eqnarray}
%%%%%%%%%%%%%%%%%%%%%%%%%%%%%%%%%%%%%%%%%%%%%%%%%%%
%\end{widetext}
where
%%%%%%%%%%%%%%%%%%%%%%%%%%%%%%%%%%%%%%%%%%%%%%%%%%%
\begin{eqnarray}
&&\Delta_{0}=J(\delta-\lambda)=\frac{1}{6}(3J_{1}-2J_{2}-J_{3}),
 \label{Bare Masses}
\end{eqnarray}
%%%%%%%%%%%%%%%%%%%%%%%%%%%%%%%%%%%%%%%%%%%%%%%%%%%
and the velocity of spin excitations
$$
u=a_{0}J(1+6\lambda) \equiv a_{0}J_{eff}.
$$

From the exact solution of the sine-Gordon
model, it is known that for
arbitrary finite $\Delta_{0}$ the excitation spectrum of the
Hamiltonian Eq.~(\ref{Bosonized_Hamiltonian}) is gapped and
consists of solitons and antisolitons with mass $M$ and
soliton-antisoliton bound states ("breathers") with the lowest
breather mass also equal to $M$ \cite{Dashen75,Takhtadjan75}. The
soliton mass $M$, which determines the gap in the excitation
spectrum and is connected with bare model parameters ($\Delta_{0}$
and $J_{eff}$) as follows $M = J_{eff}{\cal
C}(\Delta_{0}/J_{eff})^{2/3}$
%%%%%%%%%%%%%%%%%%%%%%%%%%%%%%%%%%%%%%%%%%%%%%%%%%%
%\be M = J_{eff}{\cal C}(\Delta_{0}/J_{eff})^{2/3}\, ,
%\label{SG-mass_Zamolodchikov} \ee
%%%%%%%%%%%%%%%%%%%%%%%%%%%%%%%%%%%%%%%%%%%%%%%%%%%
where ${\cal C} \simeq 1.4$ \cite{Zamolodchikov95}.

The gapped character of the excitation spectrum results to
suppression fluctuations in the system and the  $\phi$ field is
condensed in one of its vacua ensuring the minimum of the
dominating potential energy

%%%%%%%%%%%%%%%%%%%%%%%%%%%%%%%%%%%%%%%%%%%%%%%%%%%%
\begin{eqnarray}  \label{Ordered-Field}
\langle\phi \rangle  = \left\{
\begin{array}{l}
\sqrt{\pi/2} \hskip0.5cm \textrm{at \quad $\Delta_{0} < 0$} \\
0 \hskip1.1cm \textrm{ at \quad $ \Delta_{0} > 0$}
\end{array}
\right. \, .
\end{eqnarray}
%%%%%%%%%%%%%%%%%%%%%%%%%%%%%%%%%%%%%%%%%%%%%%%%%%%%%%%%%%%%%%%%%%%%%%
The vacuum expectation value of the {\em cosine} field in the
gapped phase in the weak coupling is given by \
%%%%%%%%%%%%%%%%%%%%%%%%%%%%%%%%%%%%%%%%%%%%%%%%%%%%%%%%%%%%%
\begin{eqnarray}
\epsilon=\langle\cos{\sqrt{2 \pi }\phi}\rangle \simeq
(M/J_{eff})^{1/2}\, ,
\end{eqnarray}
%%%%%%%%%%%%%%%%%%%%%%%%%%%%%%%%%%%%%%%%%%%%%%%%%%%%%%%%%%%%%%
while in the strong coupling, at $|M| \geq J$ it becomes of the
unit order \cite{Luk_Zam_97}.

In absence of the hexamer distortion, at $\lambda=0$ (i.e.
$J_{2}=J_{3}=J(1-\delta)$), $\Delta_{0}= \delta J>0$, the
excitation spectrum is gapped and the field $\phi(x)$ is pinned
with vacuum expectation value $\langle\phi \rangle  =0$. Using the
bosonized expressions for spin operators one can easily get the
on-site spin order is strongly suppressed, while the link-located
dimer order
%%%%%%%%%%%%%%%%%%%%%%%%%%%%%%%
\begin{eqnarray}
 \label{D_r_1}
\langle\textbf{S}_{n}\cdot\textbf{S}_{n+1}\rangle &=&
%Const +
%\frac{(-1)^{n}}{\pi^{2}}\langle\cos\sqrt{2\pi}\phi\rangle
%\nonumber\\&=&
Const + (-1)^{n}\frac{\epsilon}{\pi^{2}}\,
\end{eqnarray}
%%%%%%%%%%%%%%%%%%%%%%%%%%%%%%%
shows the long-range ordered for
dimer order and the maxima of the
dimerization functions located on even links.

At $\delta=0$, i.e. $J_{1}=J_{2}=J$, $J_{3}=J(1+6\lambda)$ again
the excitation spectrum is gapped, but in this case with the bare
mass $\Delta_{0}= -\lambda J<0$. In this case  for $\lambda<0$,
$\phi(x)$ is pinned with vacuum expectation value $\langle\phi
\rangle  =0$ and for $\lambda>0$,  $\phi(x)$ is pinned with vacuum
expectation value $\langle\phi \rangle  =\pi$. As a results,
fluctuations of the on-site degrees of freedom are fully
suppressed , while the link-located dimerization function shows a
long range order
%%%%%%%%%%%%%%%%%%%%%%%%%%%%%%%
\begin{eqnarray}
 \label{C-K} \langle\textbf{S}_{n}\cdot\textbf{S}_{n+1}\rangle
= Const +(-1)^{n}\,sign(\lambda)\, \frac{\epsilon}{\pi^{2}}.
\end{eqnarray}
%%%%%%%%%%%%%%%%%%%%%%%%%%%%%%%
At $\lambda<0$, exchange on each 6-th weak link becomes weaker,
the dimer order introduced by the hexamer distortion of the spin
exchange is once again given by Eq. (\ref{D_r_1}) and therefore as
in the case with $\delta > 0$  maxima of the dimerization
functions are located on even links.

At $\lambda>0$ the dimer order is pinned with the strongest links
and the minimum of the energy is realized with dimer order which
shows maxima on even, including the strongest each 6-th link.
Therefore, in  marked contrast with $\lambda<0$ case, at
$\lambda>0$ the hexamer distortion plays the role of frustration
with respect to initial explicit
dimer order.

Therefore, at finite $\delta>0$ and $\lambda>0$, in the gapped
phase, the long range ordered dimerization pattern has maxima on
odd links at $\Delta_{0}>0$ and on even links, at $\Delta_{0}<0$.
In the former case, the distribution of dimers coincides with the
order of non-disturbed dimerized chain, while in the case of
$\delta_{0}=0$, the order determined with dominant strong links.
These two gapped phases with shifted on one lattice unit with
respect to each-other pattern of dimers are separated by the
critical line where $\Delta_{0}=0$ i.e.
%%%%%%%%%%%%%%%%%%%%%%%%%%%%%%%%%%%%%%%%%%%%%%%%%%%
\begin{eqnarray}
3J_{1}-2J_{2}-J_{3}&=& 6 J(\delta-\lambda)=0\, ,
 \label{Bare Masses}
\end{eqnarray}
%%%%%%%%%%%%%%%%%%%%%%%%%%%%%%%%%%%%%%%%%%%%%%%%%%%
 and the gapless critical phase which described by the gaussian
free field is realized.

In order to investigate the detailed behavior of the ground state
 phase diagram and to test the validity of the picture
obtained from continuum-limit bosonization treatment, below in
this paper we present results of numerical calculations using the
 exact diagonalization for finite chains.
 %%%%%%%%%%%%%%%%%%%%%%%%%%%%%%%%%%%%%%%%%%%%%%%%%************FIG 2************%%%%%%%%%%%%%%%%%%%%%%%
\begin{figure}[h]
\includegraphics[width=15pc]{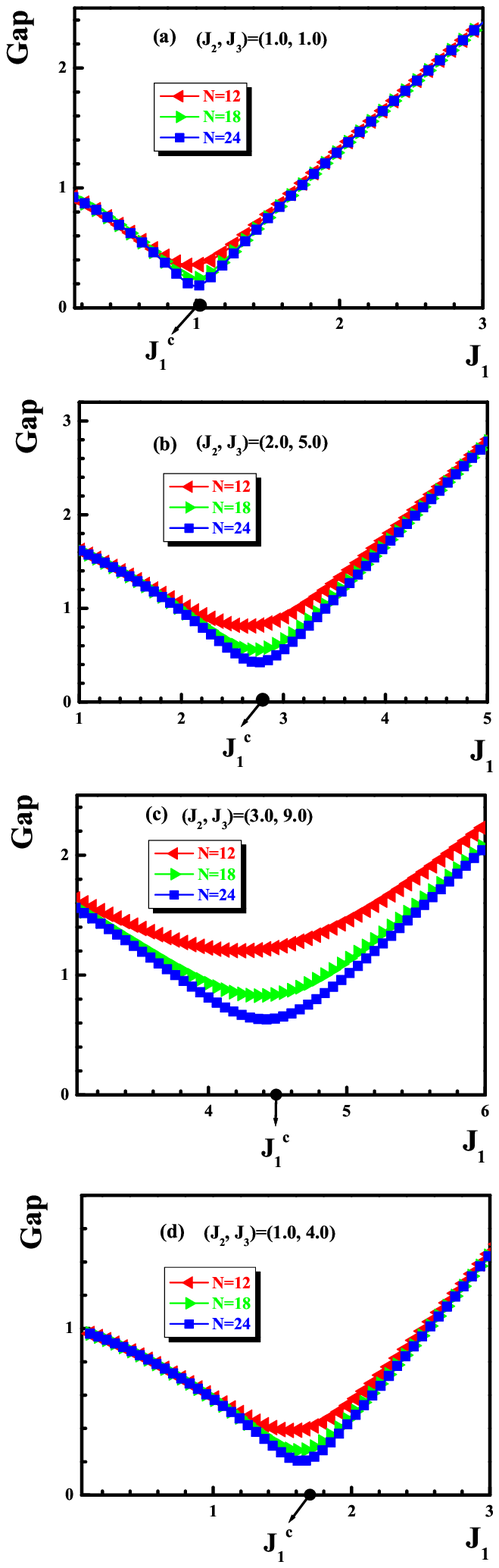}
 \caption{(Color
online) The energy gap versus $J_{1}$ for the dimerized chains
with lengths $N=12, 18, 24$ and different values of the exchanges
$(J_{2},J_{3})=$~$(a):(1.0,1.0)$,~$(b):(2.0,5.0)$ ,
$(c):(3.0,9.0)$ and $(d):(1.0,4.0)$.  }\label{Gap}
\end{figure}
%%%%%%%%%%%%%%%%%%%%%%%%%%%%%%%%%%%%%%%%%%%%%%%%%%%%%%%%%%%%%%%%%%%%%%%%%%%%%%%%%%%%%%%%%%%%%%%%%

 %%%%%%%%%%%%%%%%%%%%%%%%%%%%%%%%%%%%%%%%%%%%%%%%%%%%%%%%%%%%%%%%%%%%%%%%%%%%%%%%%%%%%%%%%%%%%%%%%%%%%%%%%%%%%%%%%%%%%

\section{Numerical experiment}

%%%%%%%%%%%%%%%%%%%%%%%%%%%%%%%%%%%%%%%%%%%%%%%%%%%%%%%%%%%%%%%
In this section, we study the ground state phase diagram of the
dimerized Heisenberg chain with spin hexameric distortion of
exchange by performing a numerical experiment. One of the remarkable ways in field of
numerical experiments is known as the exact diagonalization (ED)
method. We have used the ED
method to diagonalize numerically the Hamiltonian
(\ref{Hamiltonian}) with periodic boundary
conditions. To determine the ground state phase diagram of the
model, we have calculated the
spin gap and the dimer order parameter as a function of coupling exchange on odd links, $J_{1}$, for finite chains with lengths up to $N=24$ and
different values of exchanges $J_{2},
J_{3}$.

 In Fig.~\ref{Gap}, we have presented results of
our numerical calculations on the energy gap for the values of the
exchanges parameter corresponding to $(J_{2},J_{3})=(a): (1.0,
1.0), (b): (2.0, 5.0), (c): (3.0, 9.0)$ and $(d): (1.0, 4.0)$ for
different chain lengths $N=12, 18, 24$. We
have recognized the energy gap in
finite chains as the difference between the energies of the ground
and first exited states. It is clearly seen that the spectrum of
the model is gapped at $J_{1}=0$. As soon as the $J_{1}$ increases
from zero,  the energy gap decreases  and vanishes at the critical
value $J_{1}^{c}$. It is obvious that for finite systems, the
energy gap is always finite and only vanishes at certain value
$J_{1}^{c}$ in the thermodynamic limit $N\rightarrow\infty$. By
further increasing of $J_{1}$,
the energy gap reopens and behaves almost linearly in respect
$J_{1}$. Therefore, the peculiar behavior of the energy gap shows
that the system can be found in two different gapped phases by
tuning the exchange values.

 At $J_1=0$, the system reduced into $N/2$ pair of spins in the singlet state.
 It is completely natural to expect for finding the  ground state of the system in a phase with dimerization on even links, so called Dimer-I.
 On the other hand, in the limit of very strong exchange on odd links, $J_2=J_3=0$, the ground state has the structure of singlets on odd links, so called Dimer-II.
  So, the competition between two
terms in the Hamiltonian should
be lead to a quantum phase
transition between these two dimer phases. It is known that
due to the nature of a dimer
phase, translation invariance
symmetry is broken by one unit cell of the lattice
\cite{Sierra_07}. The structure of the dimer phase can be obtained
by studying the dimer order parameter defined as
%%%%%%%%%%%%%%%%%%%%%%%%%%%%%%%%%%%%%%%%%%%%%%%%%%%%%%%%%%
\begin{eqnarray}
D=\frac{2}{N}\sum_{n=1}^{N/2}<Gs|\textbf{S}_{2n-1}.\textbf{S}_{2n}-\textbf{S}_{2n}.\textbf{S}_{2n+1}|Gs>,
 \label{dimer}
\end{eqnarray}
%%%%%%%%%%%%%%%%%%%%%%%%%%%%%%%%%%%%%%%%%%%%%%%%%%%%%%%%%%
where the notation $\langle Gs|...|Gs\rangle$ represents the
ground state expectation value. It is expected that the dimer
order parameter, $D$, takes the values $\frac{3}{4}$ and
$\frac{-3}{4}$ in the saturated Dimer-I and Dimer-II phases
respectively. %%%%%%%%%%%%%%%%%%%%%%%%%%%%%%%%%%%%%%%%%%%%%%%%%%
%%%%%%%%%%%*%%%%%%%%%%%%%%%%%%************FIG 3********************%%%%%%%%%%%%%%%%%%%%%%%%%%%%%%%
\begin{figure}
\includegraphics[width=16pc]{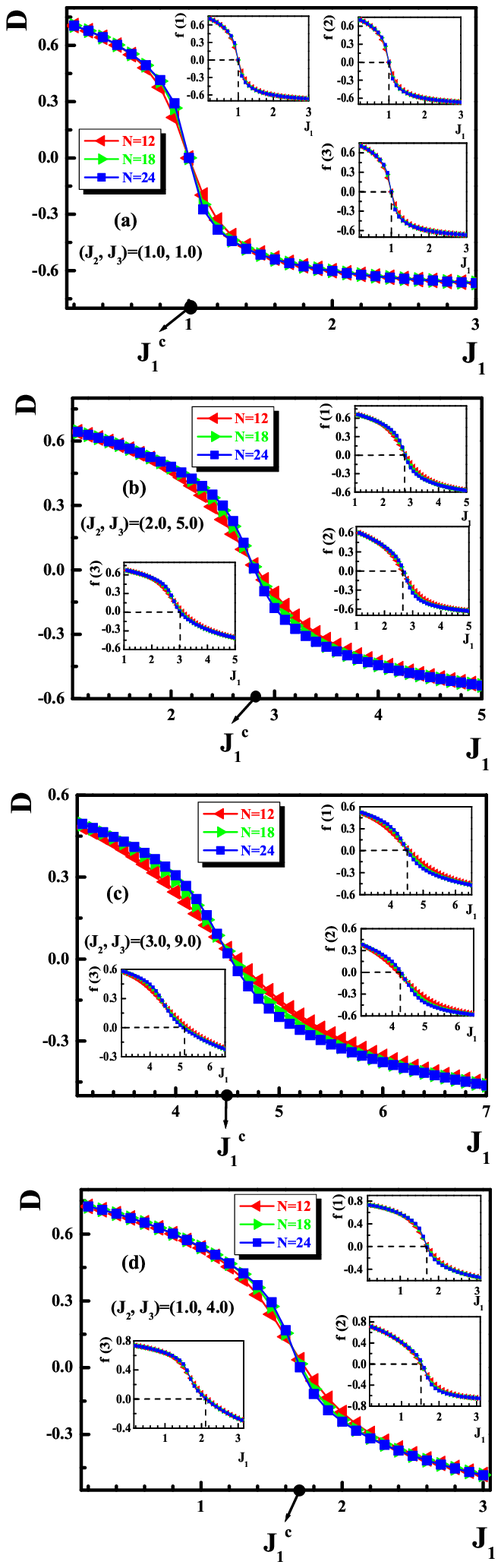}
\caption{(Color online) The dimer order parameter versus $J_{1}$
for the dimerized chains with lengths $N=12, 18, 24$ and different
values of the exchanges
$(J_{2},J_{3})=$~$(a):(1.0,1.0)$,~$(b):(2.0,5.0)$ ,
$(c):(3.0,9.0)$ and $(d):(1.0,4.0)$. In the inset, the functions
$f(1), f(2)$ and $f(3)$ are plotted. The dashed line is only a
guide for the eye. }\label{Dimer}
\end{figure}
%%%%%%%%%%%%%%%%%%%%%%%%%%%%%%%%%%%%%%%%%%%%%%%%%%%%%%%%%%%%%%%%%%
We have presented our numerical
results on the dimer order parameter in Fig.~\ref{Dimer} for
finite chains with lengths $N=12, 18, 24$ and exchange values
$(J_{2}, J_{3})=(a): (1.0, 1.0), (b): (2.0, 5.0), (c): (3.0, 9.0)$
and $(d): (1.0, 4.0)$. As is clearly seen, in the absence of
$J_1$, the dimer order parameter is positive and the ground state
of the system is in the saturated Dimer-I phase. As soon as the
$J_1$ is added and
increased from zero, induced
quantum fluctuations reduce
the dimer order parameter $D$ and it reaches to zero value at the
critical point $J_{1}^{c}$. We
 found that, taking zero of
$D$ at the critical value $J_{1}^{c}$ is size-independent which
shows its validity in the thermodynamic limit $N\longrightarrow
\infty$. By more increasing  $J_{1}$, the dimer order parameter
becomes negative which is the indication of the long-range  dimer
ordering on odd links, namely Dimer-II phase. In this region, the
negative value of $D$ enhances by increasing the exchange $J_{1}$
and it tends to the saturation value of the Dimer-II phase.
%%%%%%%%%%%%%%%%%%%%%%%%%%%%%%%%%%%%%%%%%%%%%%%%%%%%%%%%%%%%%%%%%%%%%%%%%%%%%%%%%%%%%%%%%%%%%%%%%%%%%%%%%%

To find a better physical picture of the gapped dimer phases, we
have a look into the microscopic
behavior of the system
by using our numerical
experiment. Since in a unit cell of a hexamer chain
(Fig.~\ref{Schematic}), there are three different links we have
calculated the dimerization on each link defined as
\begin{eqnarray}
f(1)&=&<\textbf{S}_{6n-5}.\textbf{S}_{6n-4}-\textbf{S}_{6n-4}.\textbf{S}_{6n-3}>,\nonumber
\\
f(2)&=&<\textbf{S}_{6n-3}.\textbf{S}_{6n-2}-\textbf{S}_{6n-2}.\textbf{S}_{6n-1}>,\nonumber
\\
f(3)&=&<\textbf{S}_{6n-1}.\textbf{S}_{6n}-\textbf{S}_{6n}.\textbf{S}_{6n+1}>.\nonumber
\\
\label{dimer}
\end{eqnarray}
%%%%%%%%%%%%%%%%%%%%%%%%%%%%%%%%%%%%%%%%%%%%%%%%%%%%%%%%%%%%%%
In the \emph{Inset} of Fig.~\ref{Dimer}, We have plotted numerical
results on these microscopic functions versus the exchange $J_1$
for chains with lengths $N=12,18,24$ and exchanges $(J_{2},
J_{3})=(a): (1.0, 1.0), (b): (2.0, 5.0), (c): (3.0, 9.0)$ and
$(d): (1.0, 4.0)$.  It is clearly seen that the values of functions
$f(1), f(2)$ and $f(3)$ take zero at different values of $J_{1}$
(dashed lines). By considering the strength of exchanges, expected that the first the dimer
on the weakest links takes zero and finally, the strongest one becomes zero
which behavior is completely seen in our numerical results.
%
%%%%%%%%%%%%%%%%%%%%%%%%%%%%%%%%%%%%%%%%%%%%%%%%%%%%%%%%%%%%%%%%%%%%%%%%%%%%%%%%%%%%%%%%%%%%%%%%%%%%%%%%%%
%%%%%%%%%%%%%%%%%%%%%%%%%%%%%%%%%%%%%%%%%%%%%%%%%%%%%%%%%%%%%%%%%%%%%%%%%%%%%%%%%%%%%%%%%%%%%%%%%%%%%%%%%%
%%%%%%%%%%%%%%%%%%%%%%%%%%%************FIG 4********************%%%%%%%%%%%%%%%%%%%%%%%%%%%%%%%%%%%%%%%%%%
\begin{figure}
\includegraphics[scale=0.7]{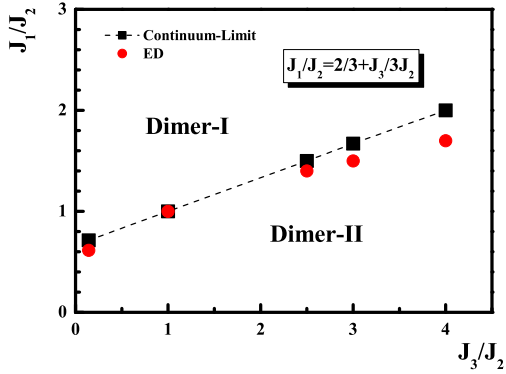}
\caption{(Color online) The phase diagram of the dimerized
Heisenberg chain with hexameric distortion of the exchange
pattern. Solid squares are the continuum-limit bosonization data,
where solid circles represents the exact diagonalization results
extrapolated into the infinite chain limit.}\label{Transition}
\end{figure}
%%%%%%%%%%%%%%%%%%%%%%%%%%%%%%%%%%%%%%%%%%%%%%%%%%%%%%%%%%%%%%%%%%%%%%%%%%%%%%%%%%%%%%%%%%%%%%%%%%
%%%%%%%%%%%*%%%%%%%%%%%%%%%%%%************FIG 5********************%%%%%%%%%%%%%%%%%%%%%%%%%%%%%%%
\begin{figure}
\includegraphics[width=20pc]{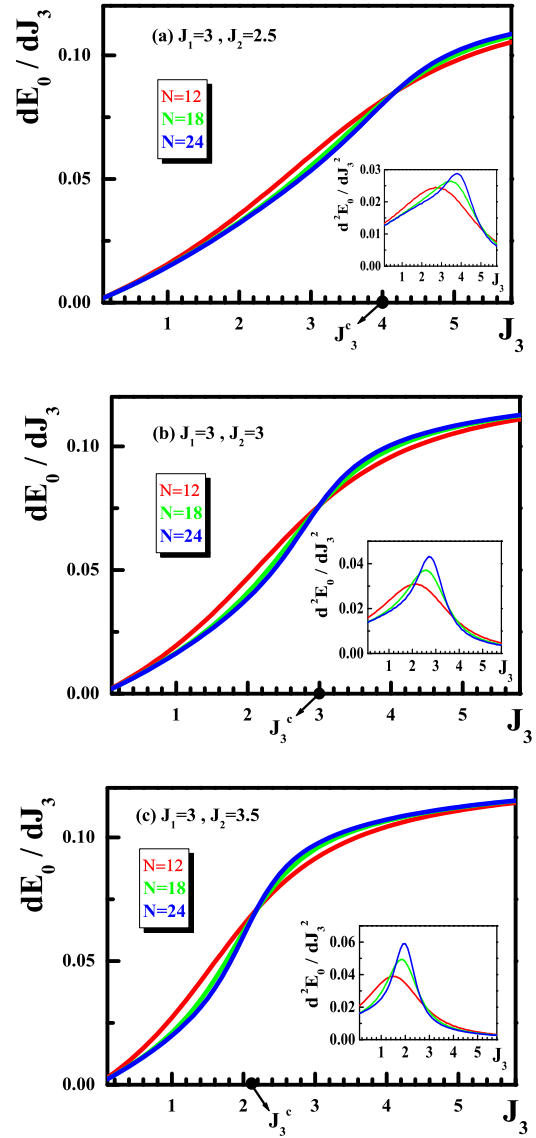}
\caption{(Color online) The first derivation of the ground state
energy as a function of the control parameter $J_3$ for the
dimerized chains with lengths $N=12, 18, 24$ and different values
of the exchanges $(J_{1},J_{2})=$~$(a):(3.0,2.5)$,~$(b):(3.0,3.0)$
and $(c):(3.0,3.5)$. In the inset, the second derivation of the
ground state energy are plotted.}\label{E-derivation}
\end{figure}
%%%%%%%%%%%%%%%%%%%%%%%%%%%%%%%%%%%%%%%%%%%%%%%%%%%%%%%%%%%%%%%%%%

We have to mention that for other values of the exchanges, we did
our numerical experiment as well and found the same ground state
magnetic phase diagram contains of: (I.) gapped Dimer-I phase in
the region $J_{1}<J_{1}^{c}$ (II.) gapped Dimer-II phase in the
region $J_{1}>J_{1}^{c}$ (Fig.~\ref{Transition}). In this figure,
solid squares are the critical points  obtained using the
continuum-limit bosonization and solid circles represents critical
points obtained by numerical experiment. The comparison of
numerical results with predictions from the continuum-limit field
theory shows well agreement.  Besides the above calculations, to
obtain the type of the mentioned quantum phase transition between
dimer phases in our model, we have implemented the algorithm to
find the ground state energy. A very important indication of the
first order phase transition is the discontinuity in the first
derivation of the ground state energy at the quantum critical
point. Using the numerical Lanczos method, we have calculated the
ground state energy for chain sizes $N=12, 18, 24$ and plotted the
first derivation of the ground state energy as a function of the
control parameter $J_3$ in the Fig.~\ref{E-derivation}. The
results of the first derivation show absence of any discontinuity
and therefore in agreement with bosonization studies indicates on
the continuous character of the phase transition. Also, we have
calculated the second derivation of ground state energy in the
inset of Fig.~\ref{E-derivation}. It is clearly seen that the
height of peak enhances by increasing $J_{3}$ and the second
derivation of ground state energy will be diverged in the
thermodynamic limit $N\longrightarrow \infty$ that is in agreement
with the continuous phase transition. The shift in the position of
peaks is result of the finite size effect.

\section{Conclusions}

In this paper, we have studied the ground-state properties of the
dimerized Heisenberg chain with spin $\frac{1}{2}$, which has
 hexameric distortion of the exchange pattern.
 We have determined the existence of a gapless line in the
 quantum phase diagram by the bosonization technique in the
 continuum-limit approach. Moreover, the exact diagonalization method has indicated that
the ground state has two gaped phases, so called the Dimer-I and
Dimer-II phases. Our numerical results are in well agreement with
analytical results.

\begin{acknowledgement}
G. I. Japaridze acknowledges support from the SCOPES Grant
IZ73Z0-128058 and the Georgian NSF Grant No. ST09/4-447.

\end{acknowledgement}

% Use the following code if you wish to generate your bibliography with BibTeX;
% replace the string "pss-demo" below with the name(s) of
% the BibTeX data base(s) you want to use.
% The resulting bibliography-output (the content of the .bbl file)
% must be pasted back into this file before submission.
% Please also include your BibTeX data base file(s) in your submission
% so that we can re-run BibTeX if necessary.
%
%\bibliographystyle{pss}
%\bibliography{pss-demo}
%
% Replace the following example bibliography with your references
% before submission:

\end{document}